\documentclass{article}

\usepackage{arvix}

\usepackage{xr}
\usepackage{bm}
\usepackage{amsmath}
\usepackage{natbib}
\usepackage{booktabs}
\usepackage[hidelinks]{hyperref}
\usepackage{graphicx}
\usepackage{caption}
\usepackage{subcaption}
\usepackage{amssymb,amsfonts,mathtools}
\usepackage{threeparttable}
\usepackage{adjustbox}
\usepackage{array}
\usepackage{pbox}
\usepackage{setspace}
\usepackage{commath}
\usepackage{enumerate}
\usepackage{xcolor}
\usepackage{ulem}
\usepackage{doi}
\usepackage{epstopdf}
\AppendGraphicsExtensions{.tiff}
\newcolumntype{P}[1]{>{\centering\arraybackslash}p{#1}}

\hypersetup{
    colorlinks=false,
    linkcolor=white,
    filecolor=black,      
    urlcolor=black
}

\def\bX{{X}}

\def\bS{{\textbf S}}

\def\bU{{U}}
\def\bS{{S}}
\def\bI{{I}}

\def\bbeta{{\beta}}

\def\bgamma{{\gamma}}

\allowdisplaybreaks

\title{Three-phase generalized raking and multiple imputation estimators to address error-prone data}

\author{
  Gustavo Amorim \\
  Department of Biostatistics \\
  Vanderbilt University Medical Center \\
  Nashvile, TN, U.S.A.\\
  \texttt{gustavo.g.amorim@vumc.org} \\
   \And
  Ran Tao \\
  Vanderbilt Genetics Institute \\
  Vanderbilt University Medical Center \\
  Nashvile, TN, U.S.A.\\
  \And
  Sarah Lotspeich \\
  Department of Biostatistics \\
  Vanderbilt University Medical Center \\
  Nashvile, TN, U.S.A.\\
  Department of Biostatistics, Gillings School of Global Public Health  \\
  University of North Carolina at Chapel Hill \\
  Chapel Hill, Chapel Hill, NC, U.S.A.\\
  \And
  Pamela A. Shaw \\
  Kaiser Permanente Washington Health Research Institute \\
  Seattle, WA, U.S.A.\\
  \And
  Thomas Lumley \\
  Department of Statistics \\
  University of Auckland \\
  Auckland, New Zealand\\
  \And
  Rena C. Patel \\
  Department of Medicine \\
  University of Washington \\
  Seattle, WA, U.S.A.\\
  \And
  Bryan E. Shepherd \\
  Department of Biostatistics \\
  Vanderbilt University Medical Center \\
  Nashvile, TN, U.S.A.\\
}

\renewcommand{\shorttitle}{Three-phase generalized raking and multiple imputation estimators to address error-prone data}

\begin{document}
\maketitle

\begin{abstract}
Validation studies are often used to obtain more reliable information in settings with error-prone data. Validated data on a subsample of subjects can be used together with error-prone data on all subjects to improve estimation. In practice, more than one round of data validation may be required, and direct application of standard approaches for combining validation data into analyses may lead to inefficient estimators since the information available from intermediate validation steps is only partially considered or even completely ignored. In this paper, we present two novel extensions of multiple imputation and generalized raking estimators that make full use of all available data. We show through simulations that incorporating information from intermediate steps can lead to substantial gains in efficiency. This work is motivated by and illustrated in a study of contraceptive effectiveness among 82,957 women living with HIV whose data were originally extracted from electronic medical records, of whom 4855 had their charts reviewed, and a subsequent 1203 also had a telephone interview to validate key study variables.
\end{abstract}



\section{Introduction}
\label{sec_introduction}

Measurement errors are often found in routinely collected clinical data. If not properly addressed, they may bias estimates, leading to incorrect conclusions.
Validation studies can be used to inform measurement error correction methods to address this  bias.
They typically consists of comparing the full medical record for a subset of patients in the electronic medical record (EMR) to the derived variables in the analytical dataset which are often extracted by automated algorithms. 
This two-step procedure is a special case of a two-phase study discussed in the statistical literature \citep{white1982two}. Traditional two-phase designs focus on collecting information that may be costly or hard to obtain for a sample of patients selected from the original dataset. These expensive variables may or may not be correlated with cheap-to-obtain or easy-to-measure variables that are available for the entire cohort. In validation studies, the error-prone variables are often good surrogates for their expensive, validated counterparts. Proper exploitation of this information can greatly boost statistical efficiency in downstream analyses \citep{tao2021efficient}. 

Sometimes, however, a single validation step (e.g., a round of chart reviews) may not be sufficient to substantially reduce measurement errors and a second validation step (e.g., follow-up telephone interviews) may be required. This was the case in our motivating example: a pregnancy study among women living with HIV using contraceptives and antiretroviral therapy (ART) \citep{patel2021pregnancies}. This study aimed to verify whether efavirenz-containing ART decreased the effectiveness of contraceptive implants. EMR data from 82,957 women were collected from 2011 to 2015. A standard analysis of the EMR data suggested that the effectiveness of implants was indeed reduced. However, concerns over data quality were raised leading to a two-stage validation study. In the first stage of validation, 4855 women were sampled from the EMR dataset to have their exposures of interest, ART regimen and contraceptive method, as well as their outcome, pregnancy status, validated through chart review. Later, a second stage of validation was conducted with a subsample of 1203 women from the chart review dataset selected for a telephone interview to further validate these variables. Those women who received the telephone interview had fully validated data whereas those whose charts were reviewed but did not have a telephone interview had partially validated data. Hence, this validation strategy resulted in a three-phase study with phase-1 being the EMR, phase-2 being the chart reviews, and phase-3 being the telephone interviews. 

Even though we focus on a specific example, three-phase studies are not uncommon. \citet{whittemore1997multi}, for example, described a genetic study of prostate cancer.
Men with and without a recent history of prostate cancer were classified as cases and controls, respectively, and basic data were recorded (phase-1). Next, a subsample) was taken, on whom more detailed information regarding their family history 
was collected (phase-2. A further subsample 
provided blood samples or archived tissue for DNA analysis (phase-3).
Another example of a three-phase study is the coronary heart disease (CHD) section of the Women's Health Initiative study, discussed in \citet{breslow2013using}. The objective was to investigate whether hormone therapy  increased the risk of CHD.
The phase-1 data consisted of women with known CHD status and phase-2 consisted of a subsample of women that had blood samples taken to check for selected biomarkers. The phase-3 data was selected using case-control sampling, for which DNA genotype was later measured. Other examples can be found in \citet{whittemore1997multistage}, \citet{zhou2016multiple}, and \citet{pfeiffer2008combining}.

Although there are several examples in the literature of three-phase designs, methods for the analysis of data resulting from these designs are lacking or restricted to specific settings. In practice, analyses are often done via methods that were developed for two-phase designs, including conditional maximum likelihood (CML) or inverse probability weighting (IPW) as in the aforementioned papers. These methods are attractive because they are fairly easy to implement and lead to consistent estimators when the missingness is due to design \citep{breslow2013using, whittemore1997multi}. However, both estimators rely only on fully observed (i.e., fully validated) subjects and their probabilities of being sampled for phase-3. They ignore data from phase-1 and phase-2, and thus are statistically inefficient. 
We are aware of two papers that have developed methods for three-phase designs in different settings. \citet{holcroft1997efficient} extended the augmented-inverse probably weighted (AIPW) estimator proposed by \citet{Robins1994} for a three-phase study with true outcome and exposure collected in the second and third phases, respectively. \citet{mcginniss2016multiple} sequentially imputed data to address multiple missing data patterns within a single study. Although there are some similarities between these studies and ours, these methods are not directly applicable to our setting where data are collected longitudinally and all variables are partially validated in phase-2 and fully validated in phase-3.

In this manuscript, we propose three estimators that make full use of all three phases of data collection to improve estimation efficiency while reducing bias. 
Specifically, we develop a generalized raking (GR) estimator, a multiple imputation (MI) estimator, and a third estimator based on both GR and MI. GR estimators improve upon the efficiency of existing IPW and CML procedures without extra modeling assumptions \citet{sarndal2003model, breslow2009improved}. GR estimators contain, asymptotically, the best AIPW estimators and are quite easy to implement with standard software \citep{Lumley2011}. GR estimators have been applied to several two-phase problems, including validation studies \citet{oh2019raking}, but to our knowledge no implementation has been proposed for three-phase problems that resemble our motivating study. The MI estimator is a sequential, multi-stage imputation approach that treats validated data as missing data for those not selected for validation \citep{cole2006multiple}, and then multiply imputes them sequentially across the multiple phases.
Finally, the estimator based on both GR and MI, similar to that applied in \citet{han2019combining}, seeks to improve the efficiency of the GR method by using the imputed dataset. The performance of our estimators, in terms of efficiency and bias, is examined via simulations modeled after the complex validation study that motivates this work. We show substantial gains in efficiency when all validation steps are properly incorporated into the analyses. Finally, using the database of \citet{patel2021pregnancies}, all proposed methods 
are used to examine whether efavirenz-containing ART reduces the effectiveness of contraceptive implants to prevent pregnancies. We show that, similar to our simulations, using information from all validations steps leads to narrower confidence intervals in real data.

\section{Motivating example}
\label{sec_data}

A retrospective analysis was conducted on a longitudinal cohort of women living with HIV, ages 15 to 45 years old, from western Kenya from January 1, 2011 to December 31, 2015. 
The main objective was to assess whether efavirenz-containing ART decreased effectiveness of contraceptive implants as measured by pregnancy. 
Data on pregnancy, ART regimen, and contraceptive use were extracted from the EMR for a total of $n_1 = 82,957$ women, of which XX reported pregnancy during the study follow up. 
Contraceptive methods were documented at each clinic visit and the leading contraceptives were categorized as 1) implants, 2) depomedroxyprogesterone acetate (DMPA), 
3) less effective family planning (LEFP) methods, 
or 4) no contraceptive method (no family planning). 
Similarly, ART regimens were recorded at each visit and leading ART regimens were categorized as one of the following: 1) efavirenz-containing ART, 2) nevirapine-containing ART, 3) protease inhibitor (PI)-containing ART, 
or 4) no ART. 
We note that there were a few other contraceptive methods and ART regimens in the EMR \citep{patel2021pregnancies}, but as they accounted for less than $4\%$ of person-years in total and were of less interest to study investigators, they were excluded from subsequent analyses.

To address potential errors in the EMR data, a three-phase sampling scheme was devised. The $n_1 = 82,957$ women with EMR data (phase-1) were stratified into 32 groups based on combinations of pregnancy status (yes/no), ART regimen (efavirenz, nevirapine, PI, and no ART), and contraceptive use (implants, DMPA, LEFP, and no contraceptive method). 
A total of $n_2 = 4855$ women were then sampled for initial validation via paper chart review (phase-2). Different sampling probabilities were defined for each stratum, with women in more clinically relevant strata at least once during their follow-up period (e.g., pregnancy while on implant) more likely to be selected for chart review than those in strata of lower priority (e.g., not on ART and not pregnant). More information on sampling probabilities can be found in \citet{patel2021pregnancies}. The chart review validation step consisted of reviewing ART regimens, types of contraceptives, pregnancy status, and all relevant dates. A total of $n_3 = 1203$ women were then further selected for telephone interviews (phase-3). Patients who had their charts reviewed (i.e., selected for phase-2) were re-stratified, based again on their EMR data, as follows: 1) contraceptive implant and pregnant, 2) contraceptive implant and not pregnant, 3) DMPA and pregnant, or 4) other. This hierarchical order from 1) to 4), with 1) being the most clinically relevant, was used to uniquely assign patients that were in more than one group during their follow-up. Patients were then sorted in terms of priority (from 1 to 4) and approached accordingly for telephone interviews; the phase-3 sample was essentially a convenience sample based on this prioritization. In the telephone interviews, ART regimens, types of contraceptives, pregnancy status, and all relevant dates were again validated. If a woman could not remember her ART regimen or type of contraceptive during the telephone interview, the information in the chart review was used.

In summary, we have EMR data on $n_1=82,957$ women (phase-1); chart review data on a subset of $n_2=4855$ women (phase-2); and telephone interview data (in addition to chart review data) on a subset of $n_3=1203$ women (phase-3). We consider those $n_3=1203$ records with both chart review and telephone interview data to be fully validated as our gold-standard \citep{patel2021pregnancies}, whereas the $n_2-n_3=3795$ records with only chart review data are partially validated. A total of 12,437 pregnancies were recorded in the EMR, from 11,265 different women, while 1116 and 309 pregnancies were recorded in the chart and telephone interview, respectively. 

Validating time-varying variables can lead to datasets that are difficult to merge. For example, a patient's EMR  could have $K_1$ time intervals, with each interval corresponding to a certain combination of ART, contraceptive, and pregnancy; her chart review could potentially have a different number of intervals (say, $K_2$), with different starting and ending dates, and so could her telephone interview (say, $K_3$). Moreover, for both chart review and telephone interview, the validation went back to January 1, 2011, even if the respective EMR data started at a later date. Thus, the follow-up time for each patient may differ between the EMR, chart review, and telephone interview datasets.  To overcome these issues, we (1) discretized time into monthly intervals and (2) restrict the follow-up time for each woman to be the intersection of her follow-up times across all the validation steps. With this discretization, the EMR, chart review, and telephone interview datasets capture the same time periods and can be described monthly. This allows us to merge and check concordance among the EMR, chart review, and telephone interview datasets for each woman at each month. 

\begin{figure}
\centering
\caption{Monthly agreements between  (a) ART regimens and (b) contraceptive use, across the three datasets: ER, chart review, and telephone interview. Note: Columns may not sum to one due to rounding. Abbreviations: LEFP = less effective family planning; MEFP = most effective family planning; FP = family planning; NVP = nevirapine; EFV = efavirenz, ART = antiretroviral treatment; PI = protease inhibitor}
\begin{subfigure}[b]{\textwidth}
   \includegraphics[width=.9\linewidth]{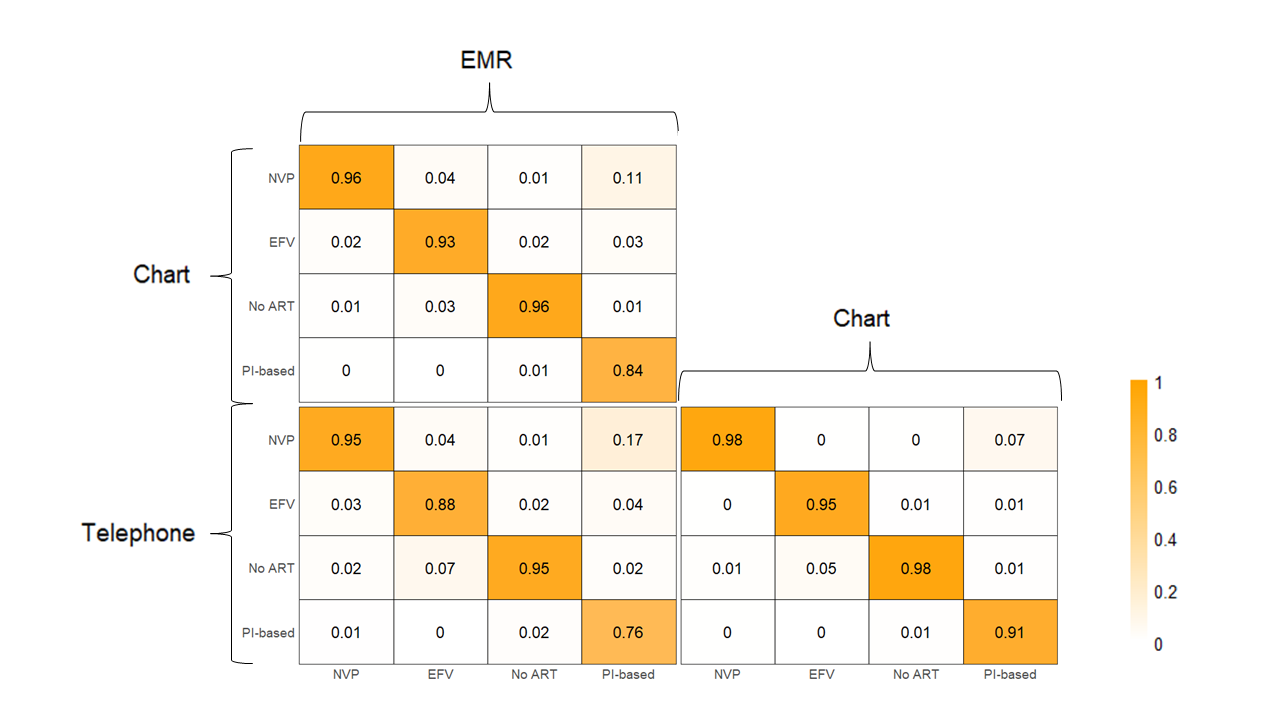}
   \caption{Monthly agreements between ART regimens recorded in the EMR and the chart review, among all women selected for chart review. Bottom half: monthly agreements between ART regimens recorded in the EMR (left side) or the chart review (right side) and ART regimens recorded in the telephone interview, among all women selected for the telephone interview.}
   \label{cross_tab_emr_phone}
\end{subfigure}

\begin{subfigure}[b]{1\textwidth}
   \includegraphics[width=.9\linewidth]{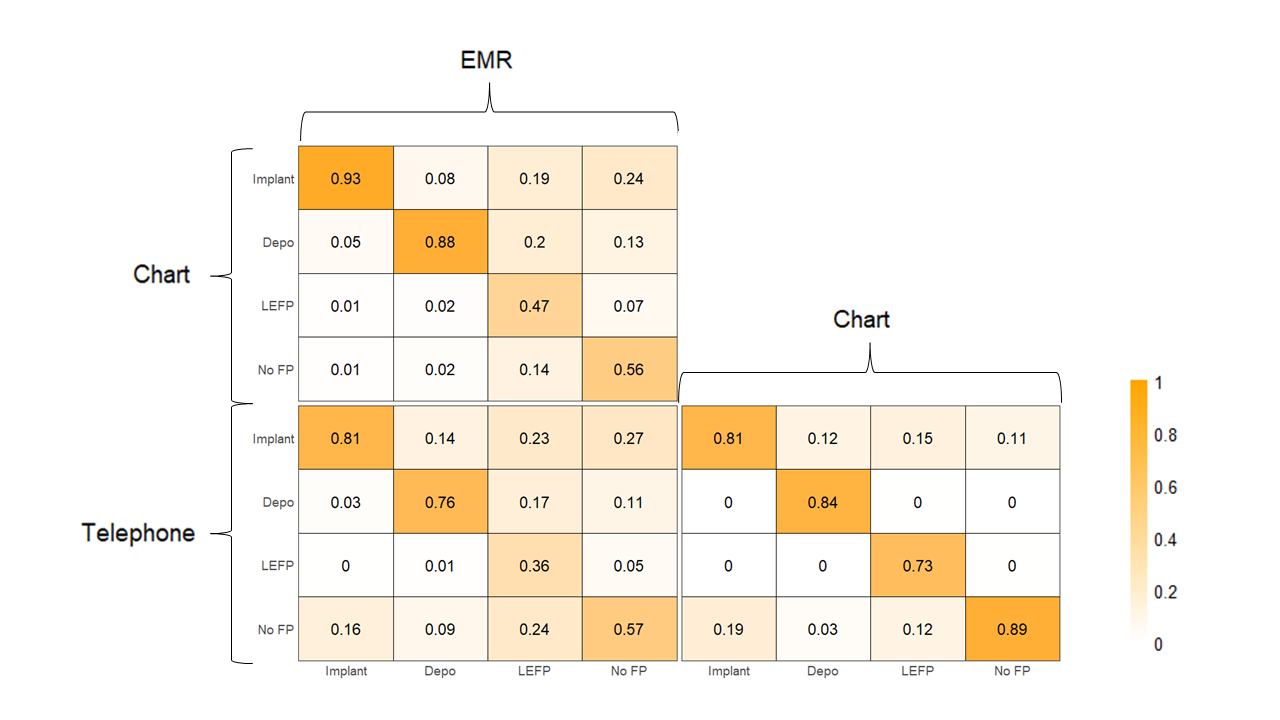}
    \caption{Monthly agreements between contraceptive/family planning types recorded in the EMR and the chart review, among all women selected for chart review. Bottom half: monthly agreements between contraceptive/family planning types registered in the EMR (left side) or the chart review (right side) and the telephone interview, among all women selected for the telephone interview.}
   \label{cross_tab_emr_phone_fptype}
\end{subfigure}

\label{fig_cross_tab}
\end{figure}

Figure \ref{fig_cross_tab} displays the monthly concordance of ART regimens and contraceptive types, respectively, across the three sampling phases. Misclassification rates were, in general, fairly small among ART regimens, especially when comparing data from the EMR to chart reviews. For example, among the $4855$ patients selected into phase-2 that had listed nevirapine-containing ART regimen as the main ART regimen in the EMR, in about $96\%$ of the months of follow up, nevirapine-containing ART was also recorded in their charts as the main ART regimen.
This concordance rate remained high when comparing the chart review to the telephone interview but decreased slightly when comparing the EMR directly to telephone interview data. In another example, among patients whose EMR listed PI-containing ART, in about 76\% of the months of follow up it was also mentioned in the telephone interviews.
In contrast, error rates in contraceptive use were substantially higher. For example, considering women selected into phase-2 with no family planning in the EMR, of the months they listed as no family planning, $24\%$ were actually found to be on implant according to the chart review. Similar error rates were seen when comparing the telephone interview and EMR datasets, with particularly high misclassification among the number of months that women whose EMR data suggested that they were on no family planning; based on telephone interviews, in about $43\%$ of such months they were said to be on some form of contraception. 

The number and time (month) of pregnancies also differed across the three datasets. Considering only women selected for chart review, 997 pregnancies were recorded in the EMR, of which only 354 were substantiated in the same month by the medical chart. Considering fully validated women, there were 180 and 237 pregnancies found in the EMR and chart review, respectively, of which only 58 and 133, respectively, were in the same months as reported in the telephone interview.

The primary goal of this study was to investigate whether efavirenz-containing ART, compared to nevirapine-containing ART, decreased the effectiveness of contraceptive implants among women living with HIV.
We calculated the incidence rate ratio (IRR) for the effectiveness of efavirenz-containing ART with contraceptive implant, compared to nevirapine-containing ART with contraceptive implant, via a Poisson regression model with pregnancy status as the outcome; contraceptive, ART regimen, as well as their interactions, as exposures of interest; and age at baseline and study site as covariates.

We start by analyzing each dataset separately, without accounting for the three-phase sampling scheme. Results are shown at the top of Table \ref{tab_pregnancy_final_rena}. All methods suggested that women on implants who were taking efavirenz-containing ART had a significantly higher risk of pregnancy than those taking  nevirapine-containing ART. However, the point estimates differed substantially, which is not surprising given the high degree of missclassification across the datasets. The IRR obtained from the telephone interview dataset was over $50\%$ higher than that using the error-prone EMR dataset only. However, these separate analyses are not only inefficient as they do not take into account all information simultaneously, but they may also be biased because they ignore the data errors and/or the sampling design. In the next section, we discuss approaches that incorporate all three datasets into a single analysis to provide valid and efficient estimates for the parameters of interest.

\begin{table}
\centering
\begin{threeparttable}
\small
\caption{Incidence rate ratio (IRR) for the effect of efavirenz-containing ART compared to nevirapine-containing ART on pregnancy, accounting for the sampling schemes.}
\small
\begin{tabular}{p{0.42\textwidth}>{\centering}p{0.15\textwidth}>{\centering}p{0.15\textwidth}>{\centering\arraybackslash}p{0.15\textwidth}}
\toprule
                       &   $\widehat{\text{IRR}}$  &  \centering \text{SD}($\log(\widehat{\text{IRR}})$) &  $95\%$ CI  \\[1ex]
\multicolumn{4}{c}{Naive analyses$\dagger$}\\[1.25ex]
EMR (phase-1 only)                          & 2.13 & \centering	0.14 & (1.63, 2.78) \\
Chart review (phase-2 only) 	            & 2.12 & \centering	0.15 & (1.58, 2.86) \\
Telephone interview (phase-3 only)          & 3.04 & \centering 0.26  & (1.82, 5.08) \\[2ex]
\multicolumn{4}{c}{Two- and three-phase analysis$\ddagger$}\\[1.25ex]
IPW                         & 3.82 &  \centering 0.52 & (1.39, 10.50) \\
two-phase GR 	        & 4.09 &  \centering 0.51 & (1.50, 11.12) \\
two-phase GR + MI       & 3.61 &  \centering 0.48 & (1.40, 9.31)  \\
three-phase GR 	        & 4.08 &  \centering 0.46 & (1.64, 10.12) \\
three-phase GR + MI     & 3.06 &  \centering 0.45 & (1.27, 7.37)  \\
two-phase MI                & 3.61 &  \centering 0.34 & (1.83, 7.11)  \\
three-phase MI              & 1.77 &  \centering 0.31 & (1.00, 3.12)  \\
\bottomrule
\end{tabular}
{\raggedright
$\dagger$: Individual analysis for each dataset (EMR, chart review, and telephone interview). Pregnancy was modelled via a Poisson regression with contraceptive type, antiretroviral therapy and their interaction as main exposures, while adjusting for age and study site (AMPATH or FACES) and using follow-up time as offset.\\
$\ddagger$: The same regression model as in the naive analysis was used, with weighting adjustments when required (IPW and GR).\\
Abbreviations: CI, confidence interval, EMR, electronic medical records; IPW, inverse probability weighting; GR, generalized raking; IRR, incidence rate ratio; MI, multiple imputation; SD, standard deviation. \par}
\label{tab_pregnancy_final_rena}
\end{threeparttable}
\end{table}

\section{Methods}
\label{sec_methods}

\subsection{Multiple imputation}
\label{sec_MI}

Validation studies are essentially missing data problems. The fully-validated data are only available for a small subsample of the study population, e.g., patients selected for telephone interview. Women with only EMRs or only selected for chart review are not fully validated, and records of these datasets are still error-prone. Multiple imputation consists of imputing the truth, i.e., correct contraceptive types, ART regimens, and pregnancy status from the telephone interview stage, multiple times for all unvalidated (phase-1, EMR data) or partially validated (phase-2, chart review data only) subjects. Conventional analyses are then performed with the complete/fully imputed data. Final estimates for the parameters of interest and their respective variances can then be computed by combining results from the multiple imputations \citep{rubin2004multiple}. MI estimates will be consistent if (1) data are missing at random (i.e., selection of records for validation is independent of the true values of the variables conditional on observed / pre-validation variables) and (2) the imputation and analysis models are properly specified.

Correct specification of the imputation models may be challenging in complex settings, for example in longitudinal studies with time-varying covariates or different follow-up times as in our motivating example. Correctly modeling the functional relationship between the time-varying covariates, such as ART regimen and contraceptive use, may be hard, and model misspecification may lead to biased estimates. We extend the time-discretized MI of
\citet{giganti2020accounting} to tackle this problem. Specifically, we deconstruct the follow-up time into time intervals, e.g., days, months, or years. We then define exposures and outcomes as present/absent in each interval. 
We decompose the joint distribution of the discretized versions of the exposure and outcome variables using chained equations. We fit these chain equations in the completely validated data and then use them to impute for unvalidated records. In a validation study framework, error-prone variables, as well as other error-free variables available for all subjects in the study can be used as covariates in the imputation procedure.

With our three-phase sampling scheme (i.e., including two validation steps), we can devise two MI procedures: 1) using the telephone interview (phase-3) to impute the missing data into the EMR dataset (phase-1) and thus ignoring the chart review (phase-2) completely (two-phase MI), or 2) first using the telephone interview to impute the missing data into the chart review dataset and then using this imputed dataset to impute data into the EMR dataset (three-phase MI). The two- and three-phase MI procedures are explained in more detail below.

Let $X_{1,m}$ and $X_{2,m}$ denote the true, fully validated ART regimen and contraceptive method, respectively, reported during the telephone interview in month $m$. Let $X^*_{j,m}$ and $\widetilde{X}_{j,m}$ denote the error-prone versions of $X_{j,m}$, for $j = 1, 2$, which are observed in the EMR (phase-1) and chart review (phase-2), respectively. Let $X_3$ be an error-free covariate or set of covariates, such as age at baseline and study site, which are available for all participants from the EMR. The variable $Y_m$ denotes the fully validated pregnancy status for month $m$ reported in the telephone interview, with $Y^*_{m}$ and $\widetilde{Y_m}$ being the error-prone outcomes observed from the EMR and chart review, respectively. 
Let $R_{1}$ be the indicator that a participant was selected for chart review and $R_{2}$ be the indicator that a participant who was selected for chart review was further selected for telephone interview; let $R=R_{1}R_{2}$.  Finally, define $O_m$ as the vector $(Y_m,X_{1,m},X_{2,m})$, $O^*_m$ as $(Y_m^*,X_{1,m}^*,X_{2,m}^*)$, and $\widetilde{O}_m$ as $(\widetilde{Y}_m,\widetilde{X}_{1,m},\widetilde{X}_{2,m})$.  Therefore, $(O^*_m,X_3,R_1,R_2)$ is observed on all subjects, $\widetilde{O}_m$ is only observed in those with $R_1=1$, and $O_m$ is only observed in those with $R_2=1$. 

The three-phase MI requires two imputation steps: imputing fully validated data for those who had chart reviews but no telephone interviews (i.e., $R_1(1-R_2)=1$) and then imputing fully validated data for those who only had EMR data ($R_1=0$). The following model is used in the first step to fit the correct pregnancy status at month $m$ for those with fully validated data ($R_2=1$): 
\begin{align}
    \label{eq_imp1}
    f_{imp}(Y_m \mid O_{m-1}^*, O_m^*, O_{m+1}^*, \widetilde{O}_{m-1}, \widetilde{O}_m, \widetilde{O}_{m+1}, X_3),
\end{align}
%
where $f_{imp}(\cdot)$ is a logistic regression model. Specific model details for \eqref{eq_imp1} and all other models are in the Web Appendix A. 
Note that the model is a function of all error-prone variables, observed in the EMR and chart review.
Fitted parameters from \eqref{eq_imp1} are then used to construct a multivariate normal distribution, from which a new set of parameters are randomly drawn, and then used to impute values, denoted by $Y^{\text{imp, chart}}_{m}$. All patients in phase-2 but not phase-3 (i.e., $R_1(1-R_2)=1$) have their true pregnancy status imputed.

An imputation model for the time-varying ART regimen $X_{1,m}$ is then constructed among those with fully validated data:
\begin{align*}
   g_{imp}(X_{1,m} \mid O_{m-1}^*, O_m^*, O_{m+1}^*, \widetilde{O}_{m-1}, \widetilde{O}_m, \widetilde{O}_{m+1}, X_3, Y_m, Y_{m-1}, Y_{m+1}),
\end{align*}
where $g_{imp}(\cdot)$ is a multinomial model. 
Correct ART regimens across months are then imputed from this model for those in phase-2 but not phase-3 in a manner similar to that for imputing pregnancy status. Note that imputed pregnancy status, $Y^{\text{imp, chart}}_{m}$, is used in place of $Y_m$ for imputation. Let the imputed ART regimens be denoted by $X^{\text{imp, chart}}_{1,m}$. 
An imputation model for $X_{2,m}$ can then be constructed for subjects with fully validated data as
\begin{align*}
    g_{imp}(X_{2,m} \mid O_{m-1}^*, O_m^*, O_{m+1}^*, \widetilde{O}_{m-1}, \widetilde{O}_m, \widetilde{O}_{m+1}, X_3, Y_{m-1}, Y_m, Y_{m+1}, X_{1,m-1}, X_{1,m}, X_{1,m+1}),
\end{align*}
and $X_{2,m}^{\text{imp, chart}}$ are imputed from this model for those in phase-2 but not phase-3 as described above. Thus, we have complete data, denoted $O^{\text{comp, chart}}_m$ on everyone in phase-2: $O_m$ for those in phase-3 and $O^{\text{imp, chart}}_m$ for those in phase-2 but not phase-3.


Similar steps are taken to impute the missing values in the EMR. Specifically, an imputation model,
\begin{equation*}
    f_{imp}(Y^{\text{comp, chart}}_{m} \mid O^*_{m-1}, O^*_m, O^*_{m+1},
    X_{3}),
\end{equation*}
is constructed among subjects in phase-2 ($R_1=1$). This model is then used to impute the correct pregnancy status for subjects only in the EMR ($R_1=0$). Let the imputed values be denoted by $Y^{\text{imp, EMR}}_{m}$. An imputation model for the time-varying ART regimen is then constructed among those with $R_1=1$ as
\begin{align*}
    g_{imp}(X^{\text{comp, chart}}_{1,m} \mid &
     Y^{\text{comp, chart}}_{m-1}, Y^{\text{comp, chart}}_{m}, Y^{\text{comp, chart}}_{m+1},
     O^*_{m}, O^*_{m-1}, O^*_{m+1}, X_3). 
\end{align*}
Monthly ART regimens, denoted $X^{\text{imp, EMR}}_{1,m}$, are then imputed from this model for those with $R_1=0$; $Y^{\text{imp, EMR}}_m$ is used in place of $Y^{\text{comp, chart}}_{m}$. Finally, an imputation model for contraceptive methods is constructed for those with $R_1=1$:
\[
    g_{imp}(X^{\text{comp, chart}}_{2,m} \mid
     Y^{\text{comp, chart}}_{m-1}, Y^{\text{comp, chart}}_{m}, Y^{\text{comp, chart}}_{m+1}, 
     X^{\text{comp, chart}}_{1,m-1}, X^{\text{comp, chart}}_{1,m}, X^{\text{comp, chart}}_{1,m+1}, 
     O^*_{m-1}, O^*_{m}, O^*_{m+1},
    X_{3}).
\]
This model is then used to impute contraceptive methods, denoted $X^{\text{imp, EMR}}_{2,m}$, for those with $R_1=0$.
%
%
The monthly complete dataset for all subjects in the EMR is then $O^{\text{comp, EMR}}_m$, which is $O_m$ for those with $R_2=1$, $O^{\text{imp, chart}}_m$ for those with $R_1(1-R_2)=0$, and $O^{\text{imp, EMR}}_m$ for those with $R_1=0$. 


Finally, a Poisson regression model
with $X^{\text{comp, EMR}}_{1,m}$, $X^{\text{comp, EMR}}_{2,m}$ and their interaction, is used to estimate the impact of efavirenz-containing ART on pregnancy, while adjusting for the error-free covariate $X_3$ and including follow-up time as an offset.
This process of imputing outcome and exposures and analyzing the fully imputed dataset is repeated $B$ times, leading to a total of $B$ estimates for the parameter of interest. Final parameter estimates are obtained by averaging them. Because the imputation and analysis models are uncongenial \citep{meng1994multiple}, an estimator for the variance is obtained using the approximation proposed by \citet{robins2000inference} and implemented in \citet{giganti2020multiple}. Notice that the three-phase MI procedure uses all error-prone variables in constructing the complete data and is thus efficient, provided that the imputation and analysis models are correctly specified.
The two-phase MI procedure is similar to that of three-phase MI except that the chart review data are ignored and the imputation is directly from phase-3 data to phase-1. Details are written out in the Web Appendix A.

\subsection{Generalized raking estimators}
\label{sec_raking}

A popular technique for handling incomplete data is the Horvitz-Thompson \citep{horvitz1952} or IPW estimator. It works by performing a weighted regression on the fully validated dataset where the weights are the inverse of the probability of being sampled for full validation.
The probability for being selected for the chart review and the probability for being further selected for the telephone interview given the chart review data are denoted by $\pi_{1i}$ and $\pi_{2i}$, respectively, such that $\pi_i = \pi_{1i} \pi_{2i}$ is the probability of subject $i = 1, \ldots, N$ being selected for full validation.
Let $\bS_i(\bbeta) = \partial \log\left\{\text{Pr}(Y_i = 1 \mid X_{1i}, X_{2i}, X_{3i}; \bbeta)\right\}/\partial \bbeta$ be the score function associated with the parameter of interest $\bbeta$. The IPW estimator, $\hat{\bbeta}_w$, is obtained by solving the estimating equation ${U}_w(\bbeta) = \sum_i {U}_{w,i}(\bbeta) = \sum_i R_id_i\bS_i(\bbeta) = {0}$ for $\bbeta$, where $d_i= 1/\pi_{i}$ is the design weight for the $i$th participant being selected for full validation. The IPW estimator assumes data are missing at random and requires that every subject has a positive probability of being sampled, i.e., $\pi_{1i} > 0$ and $\pi_{2i} > 0$ for all $i = 1, \ldots, N$. Under these assumptions, the resulting IPW estimator is asymptotically normal with variance estimated by
$\mathcal{\bI}(\hat{\bbeta}_w)^{-1}
\left\{\sum_{k,l} \bU_{w,k}^t(\hat{\bbeta}_w)\bU_{w,l}(\hat{\bbeta}_w) \right\}
\mathcal{\bI}(\hat{\bbeta}_w)^{-1}$, where $\mathcal{\bI}(\hat{\bbeta}_w) = \partial \bU_{w}(\hat{\bbeta}_w)/\partial \hat{\bbeta}_w$,

Although consistent, the IPW estimator is inefficient because it does not make full use of the dataset. It discards information on any subject that has not been fully validated. 
More efficient methods that make better use of data have been proposed for two-phase sampling, e.g., generalized rakings (GR) estimators, and they can be extended to settings with three-phase sampling to incorporate partially validated data.



GR estimators are robust approaches that build upon IPW \citep{deville1992calibration}. They use auxiliary variables available in all records to calibrate the design weights $d_i$ so that the weighted total of the auxiliary variable in the validation sample (e.g., phase-3 sample) equals the known total of the auxiliary variable in the larger sample (e.g., phase-1 sample). This is called the {\it calibration equation} and is written $\sum_i a_i = \sum_i R_i w_i a_i$, where $a_i$ is the auxiliary variable, $w_i = g_id_i$, and $g_i$ is obtained by minimizing the distance between the calibrated and design weights, $D(w_i, d_i)$, for some distance function $D(\cdot)$.

GR estimators of $\bbeta$ are more efficient than IPW estimators, while still benefiting from the same robustness properties \citep{Lumley2010}.
This gain in efficiency is obtained by carefully selecting the auxiliary variable used in the calibration equations. The goal is to find an auxiliary variable that is correlated with the true, unobserved influence function for $\beta$: higher correlation leads to higher efficiency. \citet{Breslow2009} showed that the expectation of the influence function obtained from the validation dataset, given error-prone records, is the optimal auxiliary variable. This expectation is typically unknown, but may be well-approximated using the influence function for $\beta$ based on the unvalidated data. 

Using the influence function as auxiliary variables helps connect GR and AIPW estimators. The two approaches, at first, seem very different: while AIPW, which is an important and popular technique in modern statistics, focuses on estimating the mean via regression modeling, GR estimators were first derived among survey statisticians as a regression estimator for a population total \citep{deville1992calibration, sarndal2003model}. However, if we instead set up the calibration problem to estimate the population total of the influence functions, one can show that GR and AIPW estimators are asymptotically equivalent \citep{Lumley2011}. GR estimators have, therefore, the same desirable and well-known properties of AIPW estimators: high statistical efficiency and double-robustness, i.e., they can consistently estimate the parameter of interest even if either the regression or the missingness model are incorrectly specified. This is particularly appealing when the data are missing by design, as in our motivating example.

When applied to three-phase studies, GR estimators as just described will lead to inefficient estimates. This is because GR estimators calibrate the design weights $d_i$ using information from the EMR, ignoring the chart review data. In this section we extend GR estimators to fully use all error-prone data.
We focus on a single component of $\bbeta$, say $\beta_k$. The auxiliary variables are the efficient influence functions associated with this parameter. Let $d_{1i} = 1/\pi_{1i}$ and $d_{2i} = 1/\pi_{2i}$ be the design weights for selecting the $i$th individual for chart review and further telephone interview, respectively. Similarly, let $w_{1i} = g_{1i}d_{1i}$ and $w_{2i} = g_{2i}d_{2i}$ denote the calibrated weights. The quantity $g_{ji}$, $j = 1,2$, is obtained by minimizing the distance between the calibrated and design weights under the revised constraints i) $a^*_{\text{total}} \equiv \sum_i a^*_{i} = \sum_i R_{1i} w_{1i} a^*_{i}$, which calibrates the design weights using the auxiliary variable observed in the EMR, and ii) $\widetilde{a}_{\text{total}} \equiv \sum_i R_{1i}\widetilde{a}_{i} = \sum_i R_i w_{2i} \widetilde{a}_{i}$, which calibrates the weights using the auxiliary variables observed in the chart review. The auxiliary variables $a^*_{i}$ and $\widetilde{a}_{i}$ correspond to the influence functions obtained from the error-prone EMR and chart review datasets, respectively. The efficient influence functions for $\bbeta$ are unknown, but a natural choice is to use the observed, error-prone values. Specifically, we use the influence function derived from the EMR and chart review datasets to approximate the true ones. For the Poisson model used in our analyses, these influence functions are
$a(\bX, Y; \bbeta) = \left\{\bX^t\mu(\bX;\bbeta)\bX\right\}^{-1}\bX^t\left\{Y - \mu(\bX;\bbeta)\right\}$, where $\bX = (X_1, X_2, X_3)$ and $\mu(\bX;\bbeta) = \exp\{\bX \bbeta + \log(\text{offset})\}$. Thus,
$a^* = a(\bX^*, Y^*; \hat{\bbeta}^*)$ and $\widetilde{a} = a(\widetilde{\bX}, \widetilde{Y}; \widehat{\widetilde{\bbeta}})$, where $\hat{\bbeta}^*$ is the the estimate of $\bbeta$ that uses only EMR data and $\widehat{\widetilde{\bbeta}}$ is the estimate of $\bbeta$ using only chart review data, but properly accounting for the sampling probabilities via IPW.
%

There are several types of distance functions, $D(w,d)$, each of which gives rise to a different calibration estimator \citep{deville1992calibration}. Consider, for example, the distance function $D(w_i, d_i) = (w_i-d_i)^2/2d_i$. Using Lagrange multipliers $\lambda_1$ and $\lambda_2$ to minimize $\sum_i D(w_{ki}, d_{ki})$, for $k = 1,2$, under constraints 1) and 2), we have $g_{1i} = 1 - \lambda_1^ta^*_i$ and $g_{2i} = 1 - \lambda_2^t\widetilde{a}_i$, where
$\lambda_1 = \left(\sum_{i=1}^{n_1} R_{1i} d_{1i} a^*_i {a_i^*}^t \right)^{-1} \left(\sum_{i=1}^{n_1} R_{1i} d_{1i} a^*_i - a^*_{\text{total}} \right)$
and
$\lambda_2 = \left(\sum_{i=1}^{n_1} R_{i} d_{i} \widetilde{a}_i {\widetilde{a}_i}^t \right)^{-1} \left(\sum_{i=1}^{n_1} R_{i} d_{i} \widetilde{a}_i - \widetilde{a}_{\text{total}} \right)$.

GR estimators are obtained by solving $\bU_{r}(\bbeta) = \sum_{i=1}^{n_1}\bU_{r,i}(\bbeta) = \sum_{i=1}^{n_1} R_i w_{1i}w_{2i} \bS_i(\bbeta) = {0}$ with respect to $\bbeta$.
%
%
The variance of the resulting GR estimator, $\hat{\bbeta}_{r}$, can be obtained via linearization \citep{Lumley2010}. Applying the delta method to $\sum_i \bU_{r}(\hat{\bbeta}_r) = {0}$, we have that
$\widehat{\text{Var}}\left(\hat{\bbeta}_r\right) \approx \mathcal{\bI}(\hat{\bbeta}_r)^{-1}
\left\{\sum_{k,l} \bU_{r,k}(\hat{\bbeta}_r)^t\bU_{r,l}(\hat{\bbeta}_r) \right\}
\mathcal{\bI}(\hat{\bbeta}_r)^{-1}$, where $\mathcal{\bI}(\hat{\bbeta}_r) = \partial \bU_{r}(\hat{\bbeta}_r)/\partial \hat{\bbeta}_r$. Asymptotic consistency and normality of this GR estimator follow from \citet{vanderVaart2000}.

It is important to notice that, even though we used the distance function $D(w, d) = (w-d)^2/2d$ to reach a closed-form expression for the variance of $\hat{\bbeta}_{r}$, the asymptotic distribution of $\hat{\bbeta}_{r}$ is independent of the choice of $D(\cdot)$ \citep{deville1992calibration, breslow2009improved}. Also, since $D(w, d) = (w-d)^2/2d$ may lead to negative weights, making its interpretation harder, for the remainder of this paper we will use the Poisson deviance $D(w,d) = w\{\log(w) - \log(d)\} + (d-w)$. This distance function guarantees nonnegative weights and leads to an estimator that is equivalent to the classical raking adjustment in the special case of discrete auxiliary variables \citep{Lumley2011}. 

\subsection{Generalized raking with multiple imputation}
\label{sec_RakingMI}

Here we introduce a final estimation approach that combines the GR and MI techniques discussed thus far. Recall that the two-phase GR estimator calibrates the product of the sampling weights using the influence functions obtained from the phase-1 error-prone data, while the 3-phase GR estimator uses the product of the calibrated weights obtained by calibrating each sampling weight individually using the influence functions obtained in the previous phases.

The GR with MI approach works similarly. However, instead of using the error-prone phase-1 and phase-2 data to calculate the influence functions, they are obtained from the average of all influence functions derived from each imputed dataset. That is, we first generate $B$ imputed datasets following the steps discussed in Section \ref{sec_MI} and calculate the influence functions associated with the parameters of interest for each imputated dataset. After taking the average influence function across all $B$ imputations, we apply the methods discussed in Section \ref{sec_raking} to estimate $\bbeta$.
The rationale is that by using a multiply imputed influence function we can estimate the expected value of the desired influence function given the observed data, leading to auxiliary variables that are closer to the optimal ones and thus resulting in estimates with narrower confidence intervals \citep{han2019combining}.

\section{Simulation}
\label{sec_simulation}
\subsection{Data Generation}




We studied the performance of our MI and generalized raking estimators with simulated data. The simulated data roughly followed our motivating example, with a binary error-prone outcome $Y$, a time-varying error-prone exposure $X_1$, and a time-varying error-free covariate $X_2$.
Data were generated for $i=1,\ldots, n_1=15,000$ subjects, for a follow-up of up to 18 months. We simulated three different values for $X_2$ over the entire follow-up, with constant periods of up to 6 months, while $X_1$ varied monthly.
Data for the $i$th patient can be written as ($Y_{ijk}, X_{1,ijk}, X_{2,ij}$), where $j = 1, 2, 3$ denotes the periods in which $X_2$ varied and $k = 1, \ldots, 6$ denotes the period of time (months) in which $X_1$ varied. The length of periods $j = 1, 2, 3$ were defined as the $\min\{t, 6\}$, where $t$ is the time to event (e.g., pregnancy). The rationale was the following: if an event occurred at that time, e.g., if the i$th$ woman becomes pregnant during the fourth follow-up month (so that $Y_{i14} = 1$), her follow-up stops and only returns when she is no longer pregnant. At this stage her $X_2$ status may have changed, so a new value is assigned to her. If we further assume that this woman did not get pregnant again during her follow-up, her data are written as ($Y_i, X_{1i}, X_{2i}$), where $Y_i = (Y_{i11}, \ldots, Y_{i14}, Y_{i21}, \ldots, Y_{i26}, Y_{i31}, \ldots, Y_{i36}$), $X_1 = (X_{1,i11}, \ldots, X_{1,i14}, X_{1,i21}, \ldots, X_{1,i26}, X_{1,i31}, \ldots, X_{1,i36})$ and $X_{2i} = (X_{2,i1}, X_{2,i2}, X_{2,i3})$. The time to event $T$ was generated from a Weibull distribution with shape parameter equal to $0.5$ and scale equal to $2.5\times10^{-3} \exp\left\{-\left(\gamma_1X_{1,ijk} + \gamma_2X_{2,ij}\right)\right\}$. The outcome $Y_{ijk}$ was defined as $0$ if no event occurred or 1 otherwise. Values of the time-varying exposure and covariate $X_{1,ijk}$ and $X_{2,ij}$, respectively, were independently drawn from a standard normal distribution. We set $\gamma_1 = \gamma_2 = 0.5$, leading to a prevalence for $Y$ of approximately $5\%$. 


   



Both the outcome $Y_{ijk}$ and covariate $X_{1,ijk}$ were assumed to be measured with errors, with $(Y^{*}_{ijk}, X_{1,ijk}^{*})$ and $(\widetilde{Y}_{ijk}, \widetilde{X}_{1,ijk})$ denoting the observed values of $(Y_{ijk}, X_{1,ijk})$ at phase-1 and phase-2, respectively. We assumed that the correctly recorded variables $(Y, X)$ were only available at phase-3. The error-prone variables $Y^{*}_{ijk}$ and $\widetilde{Y}_{ijk}$ were generated from separate Bernoulli distributions $\text{Pr}(Y_{ijk}^* = 1 \mid Y_{ijk}) = \text{exp}\{-3.5 + 3.25Y_{ijk}\}$ and $\text{Pr}(\widetilde{Y}_{ijk} = 1 \mid Y_{ijk}) = \text{exp}\{-5 + 4.75Y_{ijk}\}$. That way, $Y_{ijk}^*$ and $\widetilde{Y}_{ijk}$ will correctly classified a case in about $78\%$ of the times, while $Y_{ijk}^*$ and $\widetilde{Y}_{ijk}$ will incorrectly identify a case in about $3\%$ and $1\%$ of the times, respectively.

The error-prone variables $X_{1,ijk}^{*}$ and $\widetilde{X}_{1,ijk}$ were equal to $X_{1,ijk}^{*} = X_{1,ijk} + U^*$ and $\widetilde{X}_{1,ijk} = X_{1,ijk} + \widetilde{U}$, where $U^*$ and $\widetilde{U}$ were generated from mean zero normal distributions with variances 1 and 0.1, respectively. Notice that this generates data such that the chart-reviewed records $\{\widetilde{Y},\widetilde{X}_1\}$ are generally closer to the truth than the EMR-data $\{Y^*,X_1^*\}$. 

Let $E^* = 1$ denote the group of patients that had at least 1 event recorded in their EMR (phase-1) during the follow-up period and $E^* = 0$ otherwise; $\widetilde{E}$ represents the same in the chart review (phase-2). A total of $n_2 = 2500$ subjects were randomly sampled for chart review, 1250 from each group defined by $E^*$, wherein $\widetilde{Y}$ and $\widetilde{X}$ were obtained. A subsample of size $n_3$ were then sampled into phase-3 (telephone interview) for further validation. We considered two simulation settings for phase-3: 1) sampling equally from strata defined by $E^{*}$ alone, or 2) sampling equally from the strata defined by the combinations of ($E^{*}, \widetilde{E}$). Notice that simulation setting 1) ignores variables validated at phase-2, using only those observed at phase-1. We ran 1000 Monte Carlo simulations with $B = 50$ imputations for the MI approaches. Estimates were obtained by fitting the logistic model $\text{logit} \{\text{Pr}(Y_{ijk} = 1) \mid X_{1,ijk}, X_{2,ij} \} = \alpha + \beta_1X_{1,ijk} + \beta_2X_{2,ij}$ on the imputed data for all MI procedures or on the weighted fully validated data for all designed-based estimators. With rare events and short follow-up periods, $\bbeta \approx \bgamma$, so the estimates obtained from the logistic model can be used for inference \citep{ngwa2016comparison}.

We computed the empirical bias, variance, and mean squared error (MSE) for the following methods: IPW, 
two-phase and three-phase GR estimators, 
two and three-phase MI,
and two and three-phase combinations of GR and MI. 

\subsection{Results}


\begin{figure}
\centering
\begin{subfigure}[a]{1\textwidth}
   \includegraphics[width=1\linewidth]{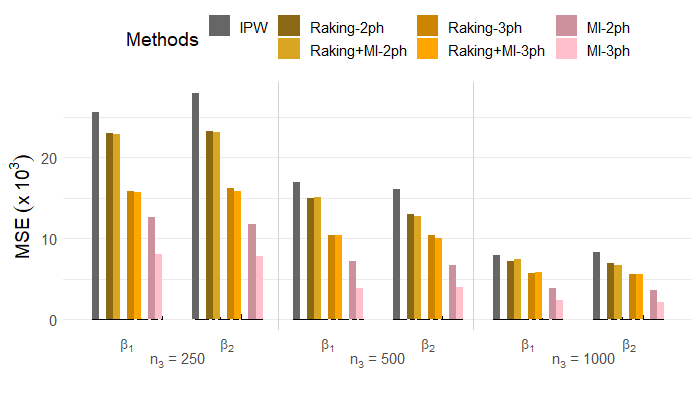}
   \caption{}
   \label{fig_3ph_setting_1} 
\end{subfigure}

\begin{subfigure}[b]{1\textwidth}
   \includegraphics[width=1\linewidth]{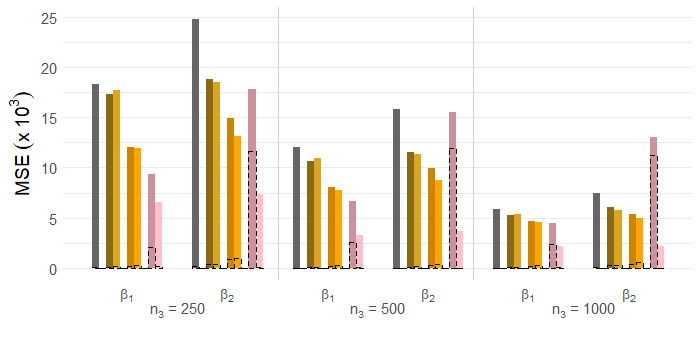}
   \caption{}
   \label{fig_3ph_setting_2}
\end{subfigure}

\caption{Mean squared error (MSE) comparing parameter estimates obtained from inverse probability weighing (IPW), two- and three-phase generalized raking (two-phase raking and three-phase raking, respectively), two- and three-phase generalized raking with multiple imputation (two-phase raking + MI and three-phase raking + MI, respectively) and two- and three-phase multiple imputation (two-phase MI and three-phase MI, respectively). (a) corresponds simulation settings 1) and (b) corresponds to simulation settings 2). Phase-3 sample size increases from left to right ($n_3 = 250, 500, \text{ or } 1000$). The dashed boxes represent the portion of the MSE due to squared bias.}
\label{fig_3ph_setting_12}
\end{figure}

Figure \ref{fig_3ph_setting_12} shows the MSE for all methods for the two simulation settings and varied phase-3 sample sizes, $n_3 = (250, 500, 1000)$. The results for empirical bias and variance are presented in Tables S1 and S2, respectively, in the Web Appendix B.
Consider first simulation setting 1) (Figure \ref{fig_3ph_setting_1}), where phase-3 sampling depended only on $E^*$. All estimators were approximately unbiased so differences in MSE reflect differences in the variance of the estimators.
As expected, the IPW estimator was least efficient, followed by the two-phase GR estimators. The three-phase GR estimators were substantially more efficient than the regular two-phase estimators. These gains in efficiency were more pronounced with smaller numbers of patients selected for phase-3, leading to estimates for $\beta_1$ that were about $30\%$ more efficient than the two-phase GR estimators and $38\%$ more efficient than the IPW estimators. 
For both two-phase and three-phase GR estimators, the MSE was similar whether weights were calibrated with the multiply imputed influence function or the naive influence function. Finally, multiple imputation with correctly specified imputation and analysis models led to even more efficient estimators. The efficiency of two-phase and three-phase MI estimators were approximately two and three times that of the IPW estimators, respectively.

Results were similar under simulation setting 2) (Figure \ref{fig_3ph_setting_2}), when the sampling probability depended on ($E^{*}, \widetilde{E}$), except for the two-phase MI estimators. The two-phase MI estimator completely ignored the phase-2 data so its missing at random assumption did not hold, resulting in a biased estimator. The three-phase MI procedure, on the other hand, correctly accounted for all sampling stages, so the final estimates were unbiased and again more efficient than all other estimators. 

\begin{figure}
    \centering
    \includegraphics[width=1\linewidth]{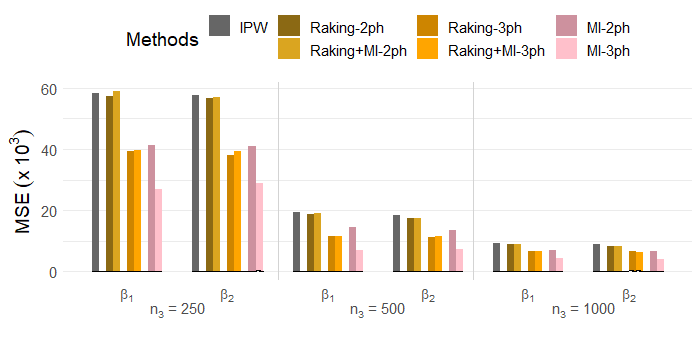}
    \caption{Mean squared error (MSE) comparing parameter estimates obtained from inverse probability weighing (IPW), two- and three-phase generalized raking (two-phase raking and three-phase raking, respectively), two- and three-phase generalized raking with multiple imputation (two-phase raking + MI and three-phase raking + MI, respectively) and two- and three-phase multiple imputation (two-phase MI and three-phase MI, respectively), under simulation setting 3. Phase-3 sample sizes increases from left to right ($n_3 = (250, 500, 1,000)$). The dashed boxes represent the proportion of the MSE due to squared bias.}
    \label{fig_3ph_bad_emr}
\end{figure}

We performed an additional set of simulations with higher phase-1 error rates. Specifically, the error-prone variable $Y_{ijk}^{*}$ was generated from a Bernoulli distribution with $\text{Pr}(Y_{ijk}^* = 1 \mid Y_{ijk}) = \text{exp}\{-2.5 +     1.75Y_{ijk}\}$, so that in about $47\%$ of the times $Y_{ijk}^*$ will correctly classify an event. Subjects were selected for phase-2 and phase-3 based only on $E^*$, as in sampling scenario i) above.  Results are displayed in Figure \ref{fig_3ph_bad_emr}, and Tables S1 and S2 of Web Appendix B. The two-phase GR and MI estimators were greatly affected by this higher error rate because the unvalidated EMR data were much less correlated with the truth and contained little information to calibrate or impute. In contrast, the three-phase generalized raking and MI estimators resulted in efficiency gains because they effectively incorporated the phase-2 chart review data which was highly correlated with the truth.

\begin{figure}
    \centering
    \includegraphics[width=1\linewidth]{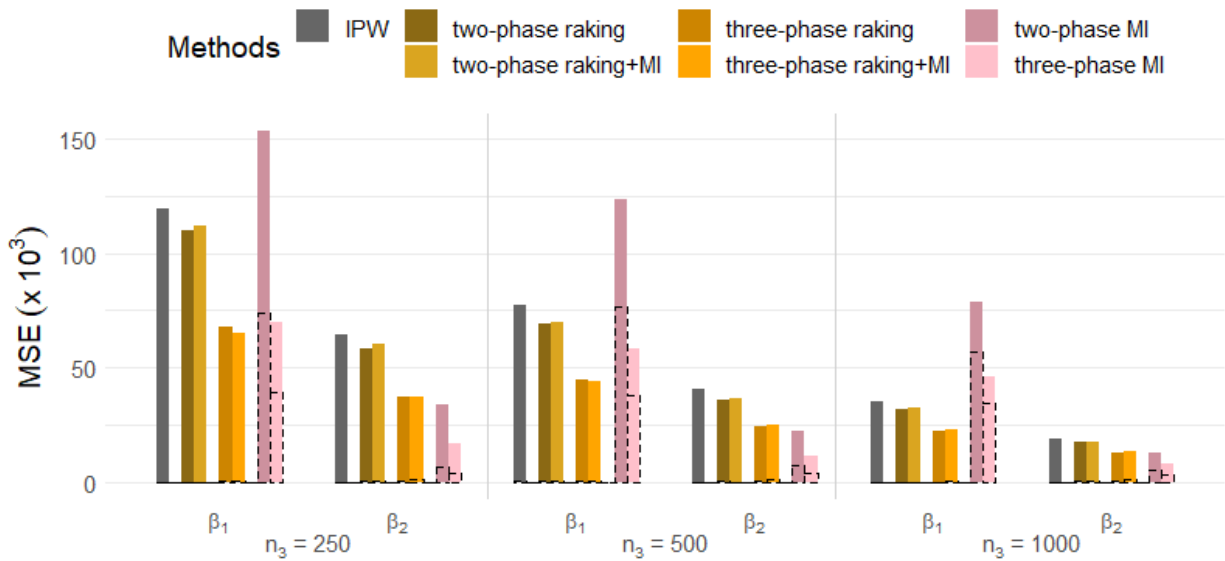}
    \caption{Mean squared error (MSE) comparing parameter estimates obtained from inverse probability weighing (IPW), two- and three-phase generalized raking (two-phase raking and two-phase raking, respectively), two- and three-phase generalized raking with multiple imputation (two-phase raking + MI and three-phase raking + MI, respectively) and two- and three-phase multiple imputation (two-phase MI and three-phase MI, respectively), under simulation simulation settings 1), but with misspecification of the imputation models. The phase-3 sample size increases from left to right ($n_3 = 250, 500, \text{ or } 1000$). The dashed boxes represent the portion of the MSE due to squared bias.}
    \label{fig_3ph_gamma}
\end{figure}

In the previous settings, all models were correctly specified, so that when the missing at random assumption was satisfied, the MI estimators were most efficient. We performed an additional simulation to investigate the impact of a misspecified imputation model. Data were generated as in simulation simulation settings 1), except that $X_{1,ijk}=\log(Z_{ijk}) + 0.5X_{2,ij} + 0.1X_{2,ij}^2$ with $Z_{ijk}$ generated from a gamma distribution with shape equal to 10 and rate equal to 1. However, a normal linear model as described in Section \ref{sec_MI}, without the quadratic term, was still used to impute the fully validated $X_1$ into phase-2 and phase-1. Details for the imputation model are given in Web Appendix C. The remaining steps followed as before, and results are displayed in Figure \ref{fig_3ph_gamma}. Both MI estimators were strongly affected by misspecification of the imputation model, leading to large bias and thus large MSE. The GR estimators combined with MI were only slightly affected, if at all, by model misspecification. The GR estimators yielded the smallest MSE when estimating $\beta_1$.

\section{Revisiting the pregnancy study}
\label{sec_case_study}

We next reevaluated the pregnancy dataset from \citet{patel2021pregnancies} described in Section \ref{sec_data}, but now applying our methods to account for both validation stages. Recall that the objective was to verify whether efavirenz-containing ART decreased the effectiveness of contraceptive implants among women with HIV. The EMR contained data from $n_1=82,957$ women living with HIV, of whom $n_2=4855$ had their medical charts reviewed to validate their pregnancy status, ART regimen, and contraceptive type. A total of $n_3=1203$ women from the chart review were also contacted by telephone to have the same variables further validated. The incidence rate ratio (IRR) for the effectiveness of efavirenz-containing ART with contraceptive implant, compared to nevirapine-containing ART and contraceptive implant as the reference level, was calculated via a Poisson regression model; pregnancy status was set as the outcome, with contraceptive type, ART regimen, as well as their interactions, as exposures of interest, while further adjusting for age at baseline and study site.

IRRs obtained for all estimators are presented in Table \ref{tab_pregnancy_final_rena}. Both two- and three-phase MI estimators used $B = 5$ imputations due to intensive computational strain. We present point estimates and the corresponding 95\% confidence intervals.
In general, all methods led to point estimates that were very similar to each other with respect to the main parameter of interest: efficacy of the contraceptive implant among women on efavirenz-containing ART compared to women on nevirapine-containing ART. Conclusions were, in general, similar to the naive analysis discussed in Section \ref{sec_data}.  All methods suggested that efavirenz is indeed associated with a reduction in efficacy of the contraceptive implant, leading to higher risks of becoming pregnant.
The IRR ranged from $3.06$ to $4.09$ for all methods, except for three-phase MI (IRR = $1.77$). The three-phase MI estimator showed a substantially lower risk of being pregnant, although still significantly higher among women on efavirenz-containing ART when compared to women on nevirapine-containing ART. This smaller risk, compared to all other estimators, may be due to misspecification of the imputation models. Recall that the three-phase MI imputes missing data in two steps: first into the chart review data and later into the EMR. If any of the models is incorrectly specified, the final estimates may be biased, as shown in the simulations.

In terms of efficiency, two-phase GR resulted in narrower variance for the parameter of interest when compared to IPW, but wider interval than the two-phase raking with MI and two-phase MI only. The methods that were constructed to use all three phases of data (EMR, chart, telephone interview) led to smaller variances compared to their two-phase counterparts.
This highlights the importance of including the extra information available in the chart review.


\section{Conclusion}



In this paper, we introduced methods for the analysis of three-phase validation studies that efficiently use all three datasets: a time-discretized multiple imputation approach, a generalized raking approach, and a final estimator that combines the two methods. 
Via simulations, we illustrated the superiority of our proposed methods over existing methods that essentially discard the intermediate, partially validated phase-2 data. 
Our three-phase estimators were substantially more efficient than those derived for two-phase studies. These methods were also used to re-analyze a large three-phase validation study of women living with HIV from western Kenya. Results were generally consistent across methods, but confidence intervals were much narrower using our new methods that incorporated data from all three phases into the analyses.

Under correct modeling assumptions, MI estimators were more efficient than GR estimators; however, MI estimators require correct specification of the imputation models to be unbiased. If proper specification feels unattainable, the less efficient three-phase GR estimator may be preferable because it is consistent under fewer assumptions and is substantially more efficient than the other robust estimators, two-phase GR and IPW.
Our generalized raking estimator can be easily implemented for different settings. It uses the {\it survey} package \citep{surveyR} available in the R software \citep{refR}. 


It is important to highlight limitations in the motivating example. Due to differences in terms of follow-up periods and timing of events and exposures between the phase-1 EMR data and both the phase-2 and phase-3 validation datasets, we needed to make some simplifications to employ our methods. In particular, 
we restricted the follow-up time for each patient to be the intersection of their follow-up times across all three datasets, which resulted in some loss of information. In addition, we discretized the follow-up into monthly intervals, which may also have led to information loss, although prior two-phase studies that have employed similar discretization have seen little information loss \citep{giganti2020accounting}. Patients selected for telephone interview were sampled based on convenience; therefore, the missing at random assumption made in all of our analyses may have been violated. Finally, throughout we have implicitly assumed that the fully validated data after the telephone interview are correct; this may not be the case, and it is possible that for some patients/variables, the EMR and/or chart review data may be correct but different from the telephone interview data.

With two-phase studies, the choice of which records to validate can have a big impact on the efficiency of results \citep{McIsaac2014, amorim2020design}. Three-phase studies are often conducted via stratified random sampling, as in our motivating example. However, it may be useful to derive optimal three-phase sampling designs that target the parameter of interest. Multi-wave designs  \citet{mcisaac2015adaptive} may be warranted, as the optimal design may depend on parameters that are unknown without preliminary validation data. We are currently investigating designs of this nature.


\section*{Acknowledgements}
This research was funded by the U.S. National Institutes of Health grants R01AI131771, U01AI069911, and K23AI120855, and by the Patient-Centered Outcomes Research Institute grant R-1609-36207. The authors would like to thank investigators in the East Africa IeDEA Consortium.

\bibliographystyle{unsrtnat}  
\bibliography{phase3Main}

\end{document}